# Construction of Differentially Private Summaries over Fully Homomorphic Encryption

Shojiro Ushiyama[1], Tsubasa Takahashi[2], Masashi Kudo[3] and Hayato Yamana[4]

[1, 3, 4] Waseda University, Tokyo, Japan
{s-ushiyama, kudoma34, yamana}@yama.info.waseda.ac.jp
[2] LINE Corporation, Tokyo, Japan
tsubasa.takahashi@linecorp.com

**Abstract.** Cloud computing has garnered attention as a platform of query processing systems. However, data privacy leakage is a critical problem. Chowdhury et al. proposed Cryptε, which executes differential privacy (DP) over encrypted data on two non-colluding semi-honest servers. Further, the DP index proposed by these authors summarizes a dataset to prevent information leakage while improving the performance. However, two problems persist: 1) the original data are decrypted to apply sorting via a garbled circuit, and 2) the added noise becomes large because the sorted data are partitioned with equal width, regardless of the data distribution. To solve these problems, we propose a new method called DP-summary that summarizes a dataset into differentially private data over a homomorphic encryption without decryption, thereby enhancing data security. Furthermore, our scheme adopts Li et al.'s data-aware and workload-aware (DAWA) algorithm for the encrypted data, thereby minimizing the noise caused by DP and reducing the errors of query responses. An experimental evaluation using torus fully homomorphic encryption (TFHE), a bit-wise fully homomorphic encryption library, confirms the applicability of the proposed method, which summarized eight 16-bit data in 12.5 h. We also confirmed that there was no accuracy degradation even after adopting TFHE along with the DAWA algorithm.

## 1 Introduction

In recent years, cloud computing has garnered significant attention as a system that facilitates query processing. However, data leakage is considered a serious problem, especially when processing sensitive data on cloud servers. Contextually, we assume three entities, namely data owners that provide original data to be analyzed, a cloud server that processes the original data, and data analysts that perform arbitrary analysis through the query responses of the cloud server. In this setting, the original data could be leaked to either or both the cloud server and data analysts. This paper aims to solve the aforenoted problems by adopting fully homomorphic encryption [1] and differential privacy [2].

Fully homomorphic encryption (FHE) [1] evaluates arbitrary functions in addition and multiplication operations over encrypted data without decryption. By adopting FHE to handle the original data on the cloud server, we can preserve the privacy of this data. Note that homomorphic encryption (HE) is a limited version of FHE that enables additions or multiplication, or an arbitrary number of additions and a limited (or few) number of multiplications over encrypted data.

Differential privacy (DP) [2] is a promising privacy-preserving technique that hinders the estimation of input data by adding noise to output data. In query processing systems, we can adopt DP to preserve the privacy of individual data from query responses by adding noise. DP gives an information theoretic privacy guarantee. However, we must trust the cloud server because the cloud server handles the original data, whose privacy must be maintained, to respond queries; this means that both the original data and the query response data are revealed to the cloud server.

Research on combining HE and DP to take advantage of both techniques to preserve both the privacy of the original data provided by data owners and the privacy of output data has gained increasing research focus since around 2015 [3]. In 2020, Chowdhury et al. [4] proposed a query processing system called Crypt$\varepsilon$ that protects original data against both two cloud servers and data analysts by combining HE and DP. The DP index that they proposed summarizes a dataset via DP to successfully bound the information leakage. However, two problems persist: 1) the original data are decrypted to apply sorting via a garbled circuit, and 2) the added noise becomes large because the sorted data are partitioned with equal width, regardless of the data distribution.

To tackle these problems, we combine FHE and DP to protect the privacy of original data against both two cloud servers and data analysts by summarizing the original data without decrypting the original data, which we call DP-summary. We construct the DP-summaries in advance from the original data over FHE, followed by decrypting them to handle query processing. Then, the cloud server processes data analysts' queries with the DP-summaries, which are plaintext, to speed up the query response time such that it is the same as that of Crypt$\varepsilon$. Since all queries are processed on the DP-summaries whose privacy is guaranteed by DP, data analysts cannot make statistical guesses about the data owners' original data even if they query many times. Moreover, we adopt a part of the data-aware and workload-aware (DAWA) algorithm proposed by Li et al. [5] over FHE to reduce the query response errors caused by DP, which is another characteristic issue of Crypt$\varepsilon$.

Our contribution is stated below:

— We combine FHE and DP to protect the privacy of original data owned by data owners against both a cloud server and data analysts. During the process, we never decrypt the original data until DP is adopted to enhance the security. Moreover, we adopt the DAWA algorithm [5] over FHE to reduce the errors of query responses while Crypt$\varepsilon$ exhibits substantial errors.

This paper is organized as follows. Section 2 summarizes preliminary information regarding HE and DP. Related work is discussed in Section 3. The details of the proposed method are presented in Section 4, followed by the experimental evaluation in Section 5. Finally, we provide conclusions and discuss future work in Section 6.

## 2  Preliminaries

### 2.1  Homomorphic Encryption

FHE [1], proposed by Gentry, enables an arbitrary number of multiplications and additions over encrypted data without decryption, whereas HE enables multiplications or additions over encrypted data, or an arbitrary number of additions and a limited (or few) number of multiplications over encrypted data. Generally, HE enables faster execution than FHE. Although we do not need to decrypt the encrypted data even during the calculation by adopting HE, we cannot execute any branch operations because we cannot know the Boolean conditions. Thus, complicated functions such as square root and trigonometric functions are difficult to implement. By contrast, the adoption of bit-wise FHE such as torus fully homomorphic encryption (TFHE) [6] enables the implementation of arbitrary functions by constructing circuits. In this study, we adopt TFHE.

### 2.2  Differential Privacy

DP [2] protects the privacy of data by adding noise. A trade off exists between the strength of privacy preservation and the usefulness of differentially private data. Specifically, while adding more noise improves the privacy preservation strength, the differentially private data has a larger deviation from the original value, and the usefulness of the differentially private data decreases.

DP has a thorough mathematical basis. It is said that a randomized mechanism $m$ satisfies $\epsilon$-DP if and only if it satisfies Definition 1, given below. Then, $\epsilon$ is called a privacy parameter and takes a real value larger than 0. The size of $\epsilon$ can be used to adjust the privacy strength. Specifically, the smaller the value of $\epsilon$, the stronger the guaranteed privacy.

*Definition 1 ($\epsilon$-differential privacy [2]). A randomized mechanism* m *satisfies differential privacy if and only if the following holds:*

$$\frac{\Pr(m(D) \in S)}{\Pr(m(D') \in S)} \leq exp(\epsilon), \tag{1}$$

*where* D *and* D′ *are any pair of databases with* $d(D, D') = 1$[1], S *is any subset of the output of the randomized mechanism, and Pr() means the probability that the event in () occurs.*

Here, a randomized mechanism, e.g., Laplace mechanism [2], is a function that adds a random value to its input value to satisfy DP. The Laplace mechanism samples noise according to the Laplace distribution with a zero mean as follow: $m_{LAP}(D) = q(D) + r$, where $q$ is a query, $r$ is sampled from $Lap\left(\frac{\Delta_q}{\epsilon}\right)$, and $\Delta_q$ is the sensitivity of query $q$.

---

[1] $d(D, D') = 1$ means that the two databases $D$ and $D'$ are exactly the same except for one record, and the rest of the records are the same.

Since the noise added by DP is sampled according to a probability distribution having a mean of zero, collecting numerous differentially private output data can reconstruct its statistical properties, i.e., attackers who collect numerous differentially private output data can guess the original data probabilistically. Thus, the upper limit on the number of times that add the randomized noise is controlled by setting a privacy budget to prevent such attacks.

## 3 Related Work

### 3.1 Combination of homomorphic encryption and differential privacy

In 2020, Chowdhury et al. [4] proposed a privacy-preserving query processing system called Cryptε by combining labeled homomorphic encryption (labHE) [7], an extension of linearly homomorphic encryption, and DP. It consists of two cloud servers, one computation server and one decryption server, as shown in Fig. 1. The system protects the privacy of original data against both the two cloud servers and data analysts by applying DP under labHE. The original Cryptε has the disadvantage of a slow query response time because it performs homomorphic operations to apply DP after receiving a query. As a countermeasure to the above disadvantage, Cryptε was amended with a differentially private index, called a DP index, to accelerate range query responses. Once the DP index is constructed, all the queries are executed with the DP index. However, building the DP index requires the decryption of original data to apply sorting via a garbled circuit, which means it is possible for the sorting result to be leaked to the computation server[2]. Besides, the added noise becomes large because the sorted data are partitioned with equal widths, regardless of the data distribution.

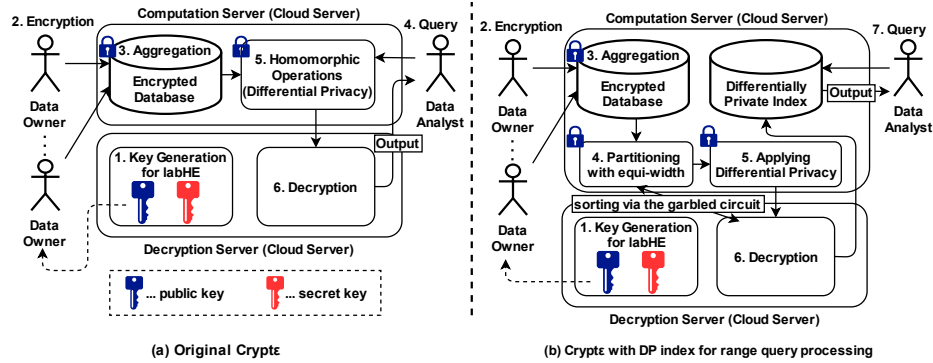

**Fig. 1.** Cryptε system [4]

---

[2] Details of Cryptε's possible privacy leakage are unknown because of no detailed implementation described in the paper [4]; thereby, some other information might be leaked.

### 3.2 Range Queries under Differential Privacy

The range query algorithm under DP proposed by Li et al. [5] in 2014 achieves low errors for range queries corresponding to one-dimensional and two-dimensional data represented by histograms. Low errors obtained via DP improve the accuracy of query responses. The algorithm consists of two steps: 1) partitioning input data represented by histograms into clusters, each of which consists of close values, and 2) optimizing the response results for a set of range queries (i.e., workload) to reduce the error. The DAWA [5] algorithm includes these two steps. Here, we call step 1) the data-aware algorithm, and step 2) the workload-aware algorithm. In the data-aware algorithm, noise is added to the total sum value corresponding to each cluster; then, the average over the cluster is calculated. Hence, this step generates two types of errors, aggregation errors by averaging and perturbation errors by adding noise. To reduce the query response errors, the data-aware algorithm seeks the best partitioning that minimizes both types of errors. Therefore, the data-aware partitioning reduces the total amount of noise rather than adding noise to the raw pieces of data. Recently, several workload-aware approaches that have outperformed the DAWA algorithm have been proposed [8]. However, our focus is constructing a differentially private summary without given workloads. Among data-aware approaches without given workloads, the DAWA algorithm is the state-of-the-art in terms of average errors [9].

## 4 Proposed Method

### 4.1 Overview

In this section, we propose a range query processing system that responds to data analysts' queries quickly, with no limit on the number of query responses, by constructing a differentially private summary (DP-summary) over FHE in advance. Specifically, our proposed method improves the query response time by responding to data analysts' queries quickly using a pre-constructed DP-summary in plaintext. Moreover, it also overcomes the limitation on the number of query responses, caused by DP, by responding to all queries from the pre-constructed DP-summary instead of applying DP to the response for each query. These advantages are the same as those of Cryptε with a DP index.

Besides, our proposed method solves two problems of Cryptε with DP index: 1) decryption of the original data before adopting DP and 2) the large amount of added noise caused by partitioning the sorted data with equi-width regardless of the data distribution in the DP index. Our proposed method solves the problem 1) by applying DP over FHE before decryption, which does not decrypt any original data until DP is adopted to enhance data security. To tackle the problem 2), we reduce the amount of added noise by adopting the data-aware algorithm, which optimizes the partitioning depending on the data distribution over FHE.

Fig. 2 shows an overview of our proposed method. We assume four entities: data owners (DOs), a computation server (CS), a decryption server (DS), and data analysts (DAs). The DOs, CS, and DS are assumed to be semi-honest; that is, they follow the protocol of our proposed system but attempt to steal the original data owned by the DOs. The DAs are assumed to be untrustworthy. It is also assumed that the CS and the DS collude neither with each other nor with other entities. Descriptions of each entity are presented below. In the following descriptions, $N$ and $M$ are arbitrary positive integers larger than or equal to 1.

- **Data Owner** ($DO\_j(\forall j, 1 \leq j \leq N)$)
  The number of DOs is $N$. DOs encrypt their own data using symmetric keys received from the DS and then send the data to the CS. After sending the data, the DOs are not involved in the system.
- **Computation Server** ($CS$)
  The CS aggregates the received encrypted data and applies DP to the encrypted data over TFHE. It also cooperates with the DS to construct the DP-summary for the received encrypted data. Any data stored on the CS are always protected by either or both TFHE and DP.
- **Decryption Server** ($DS$)
  The DS has two roles: key generation for TFHE and decryption of the DP-enabled encrypted data received from the CS. Any data decrypted by the DS is always differentially private; thus, the original data owned by the DOs is protected.
- **Data Analyst** ($DA\_i(\forall i, 1 \leq i \leq M)$)
  The number of DAs is $M$. The DAs query the CS and obtain responses to the queries from the CS. DAs can attempt to make statistical guesses regarding the DOs' original data by receiving many query responses.

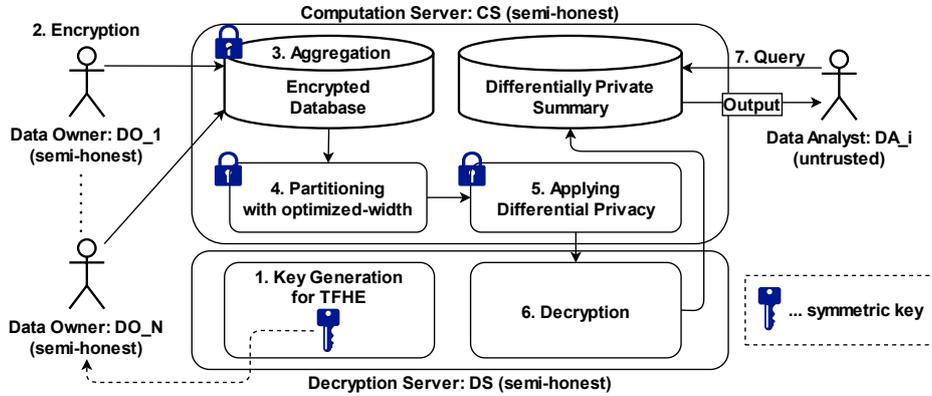

**Fig. 2.** Proposed System

Our proposed system protects the privacy of original data owned by the DOs against the CS, DS, and DAs. Further, the CS and DS are allowed to store the data secured via DP. By guaranteeing DP for the partitioning result, the CS and DS are allowed to store the partitioning result in plaintext. The symmetric keys are held only by the DOs and DS. Encryption is performed only by the DOs, and decryption is performed only by the DS.

The procedure of our proposed method is presented below based on Fig. 2. Among the following steps, the preprocessing steps from "1. **Key Generation**" to "6. **Decryption**" are completed before the DAs' queries are input.

1. **Key Generation**: The DS generates a symmetric key via TFHE. The generated symmetric key is sent securely to the DOs.
2. **Encryption**: The DOs encrypt their data using the symmetric key received from the DS and send the encrypted data to the CS.
3. **Aggregation**: After receiving the encrypted data from the DOs, the CS aggregates multiple encrypted data points whose domain is the same over homomorphic operations to construct a histogram for each pre-defined domain.
4. **Partitioning**: The CS partitions the aggregated encrypted data using homomorphic operations without decryption.
5. **Applying Differential Privacy**: The CS adds noise to each partitioned cluster using homomorphic operations and the Laplace mechanism without decryption followed by sending the differentially private encrypted data (DP-summary) to the DS.
6. **Decryption**: The DS decrypts the encrypted DP-summary received from the CS. The decrypted DP-summary is sent to the CS.
7. **Query**: The DAs query the CS and obtain the query response result.

### 4.2 Adoption of Differential Privacy over Fully Homomorphic Encryption

In this section, we explain how to adopt DP to a range query processing system based on FHE. Here, we only target range queries for histogram data; thus, we adopt the data-aware algorithm, a part of the DAWA algorithm proposed by Li et al. [5]. We consider only one-dimensional data in our present implementation. The data-aware algorithm performs the partitioning such that the sum of the deviations for each cluster is minimized. To perform the partitioning over ciphertexts, we need to calculate absolute values and minimum values over ciphertexts. Thus, we adopt TFHE [6] as an FHE scheme, specifically, as a bit-wise fully homomorphic encryption scheme. By adopting TFHE, arbitrary logic circuits consisting of binary gates can be constructed for the encrypted data to calculate absolute values and minimum values over ciphertexts.

In the data-aware algorithm, the aggregated histogram data, which is the aggregation of input data sent from the DOs to the CS, is partitioned into a set of clusters, each of which has close values such that the deviation in the cluster is small; in other words, there are approximately uniform histogram data in each cluster. Then, the noise is added to the total sum value of each cluster. Rather than adding noise to each data point, adding noise to each partitioned cluster reduces the total amount of noise added. Since partitioning the data to minimize the deviation within each cluster would result in a privacy violation, the data-aware algorithm consumes privacy budget $\epsilon_1$ to perform differentially private partitioning. Assuming that privacy budget $\epsilon_2$ is entailed when adding noise to the total sum value of the data in each partitioned cluster, the data-aware algorithm satisfies $\epsilon$-differential privacy, where $\epsilon = \epsilon_1 + \epsilon_2$.

Here, we call each histogram data a *domain*, while a cluster of domains classified by partitioning is called a *bucket*. A set of buckets is called a *partition*. We assume a set of histogram data represented by $x = (x_1, x_2, \ldots, x_i, \ldots, x_n)$, where $1 \leq i \leq n$, $n$ is the number of domains, and $x_i$ represents the data of the $i$-th domain. A set of buckets is defined as $B = (b_1, b_2, \ldots, b_j, \ldots, b_k)$, where $1 \leq j \leq k \leq n$ and $b_j$ represents the $j$-th bucket. For example, if $b_j$ is a set of domains from the third to the sixth domains, it is expressed as $b_j = \{3, 4, 5, 6\}$. $B$ is calculated from $x$ and $\epsilon_1$ in the way that minimizes the total cost of buckets; specifically, the cost of each bucket is the deviation of the data in the bucket, same as algorithm 1 of DAWA [6]. Noise generated by the Laplace mechanism by consuming $\epsilon_1$ is added to each bucket's deviation to calculate its cost; then, the final $B$ is chosen to minimize the total cost, making $B$ differentially private. Further, we define a set of the total sum values of each bucket as $S = (s_1, s_2, \ldots, s_j, \ldots, s_k)$, where $s_j$ is the sum value over the bucket $b_j$ and $1 \leq j \leq k$. To make $S$ differentially private, the CS adds noise generated from the Laplace mechanism by consuming $\epsilon_2$ to $S$. The differentially private total sum values of a given partition are defined as $S' = (s'_1, s'_2, \ldots, s'_j, \ldots, s'_k)$, where $1 \leq j \leq k$ and $s'_j$ represents the differentially private total sum value of the histogram data in the $j$-th bucket.

To deploy a DP-summary, the CS sends $S'$ to the DS to decrypt $S'$ using the symmetric key of TFHE, and then the DS sends it back to the CS as plaintext. The CS applies uniform expansion to the decrypted $S'$, i.e., dividing $s'_j$ by the number of elements in $b_j$. For example, if $s'_j = 10$ and $b_j = \{3, 4, 5, 6\}$, the uniform histogram value becomes 2.5 so that $(x'_3, x'_4, x'_5, x'_6) = (2.5, 2.5, 2.5, 2.5)$. Finally, the CS obtains the uniformly expanded data $x' = (x'_1, x'_2, \ldots, x'_i, \ldots, x'_n)$. We use $x'$ as the DP-summary to respond to DAs' queries. Note that $x$ and $S$ are represented in ciphertext and $B$ and $x'$ are represented in plaintext. $S'$ is represented in ciphertext until it is decrypted by the DS.

Fig. 3 shows an example with $x = (E(3), E(2), E(6), E(5), E(6), E(3), E(4))$ and $B = \{\{1, 2\}, \{3, 4, 5\}, \{6, 7\}\}$ as a calculated partition, where E is the encryption algorithm, i.e., TFHE. In this example, $S = (E(5), E(17), E(7))$. The differentially private total sum value per bucket is assumed to be $S' = (E(4.6), E(16.2), E(7.8))$; then, the values of the uniformly expanded data are $x' = (2.3, 2.3, 5.4, 5.4, 5.4, 3.9, 3.9)$.

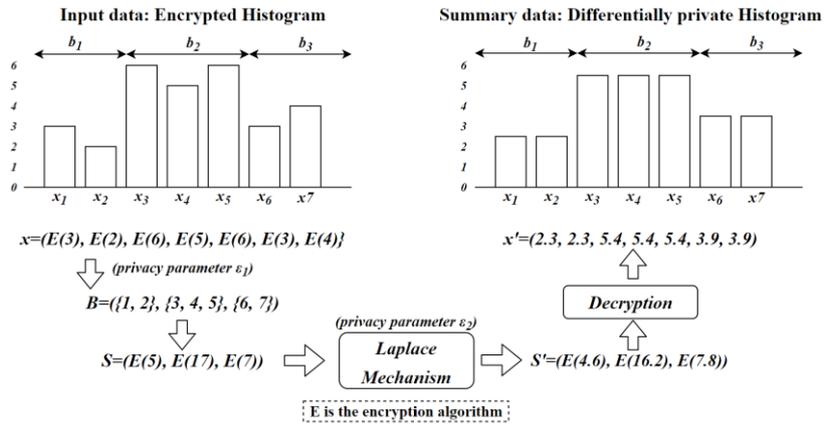

**Fig. 3.** Construction of Differentially Private Histogram over Fully Homomorphic Encryption

### 4.3 Security Analysis

The security assumption of our proposed method is presented below.

- The DOs, CS, and DS are assumed to be semi-honest; that is, they follow the protocol of our proposed system but attempt to steal the original data owned by the DOs.
- The DAs are assumed to be untrusted.
- The CS and the DS collude neither with each other nor with other entities.
- The symmetric keys are held only by the DOs and DS. (Encryption is performed only by the DOs, and decryption is performed only by the DS.)

Our proposed system protects the privacy of original data owned by the DOs against the CS, DS, and DAs. The original data are encrypted at the DO to be sent to the CS, so that the CS cannot see the original data. Then, the CS executes partitioning over the encrypted original data followed by applying differential privacy without decryption, which guarantees the CS cannot see any information related to the original data. After the DS receives the differentially private encrypted partitioned data consisting of B and S', the DS can decrypt them, which also guarantees that the DS only know differentially private data. Thus, any information related to the original data does not reveal to any parties under the condition where CS and DS never collude each other.

## 5 Experimental Evaluation

In the experimental evaluation, we examined the execution time to construct the DP-summary and its accuracy.

### 5.1 Experimental Setup

The programs used in the evaluation were written in C++ with TFHE [6] version 1.1 and run with single-threaded execution in the environment presented in Table 1. We adopted fixed-point number representation and two's complement arithmetic. In the implementation using fixed-point arithmetic, the value of the fractional part that cannot be expressed was truncated. Although approximate arithmetic (CKKS) [10] enables arbitrary polynomial functions over encrypted complex-number vectors for handling real numbers, we cannot execute branch operations such as greater-than without decryption, resulting in no partitioning. Thus, we adopt TFHE.

The privacy parameter, $\epsilon$, used in the evaluation experiment was 1.00. The ratio of $\epsilon_1$ and $\epsilon_2$ was 1:3, same as that used by Li et al. [5], i.e., $\epsilon_1 = 0.25$ and $\epsilon_2 = 0.75$. The histogram data $x_i$, $1 \leq i \leq n$ used in the experiment was generated randomly between 0 and 10. The upper limit of the data was determined to ensure that no overflow occurs during the computation process. Since negative numbers are not assumed as the numerical data, if the differentially private numerical data becomes negative, the numerical data is replaced with 0.

Table 1. Experimental Environment

| Name | Value |
| --- | --- |
| CPU model | Intel(R) Xeon(R) Platinum 8280 |
| Socket | 2 |
| Core | 56 |
| Memory size | 1.5TB |
| OS | CentOS Linux release 7.6.1810(Core) |
| Linux version | 3.10.0-957.21.3 |
| g++ version | 7.3.1 |

## 5.2 DP-summary Construction Time

To validate our proposed method's applicability, we examined the DP-summary construction time, which is the execution time taken to construct a differentially private database from the encrypted database with TFHE. The construction time depends on the domain size and the number of bits representing ciphertexts in TFHE. Thus, we changed the domain size and the bit size representing the ciphertexts to measure the construction time.

The construction time was measured from the beginning of partitioning to the end of the uniform expansion 10 times to determine the average by using the *chrono* function, which is included in C++ standard library. We changed the domain size from 2 to 8 and the bit sizes as 10(2) bits, 12(4) bits, and 16(8) bits, representing the total bit size (the bit size in the fractional part). For example, 10(2) shows 10 bits in total, in which 2 bits are used for the fractional part, 7 bits are used for the integer part, and the remaining 1 bit is used for the code part.

Fig. 4 shows the construction time based on different domain sizes; this confirms that the DP-summary construction time increases exponentially with the domain size because the number of domain combinations to merge increases exponentially to identify the best one. However, the proposed method is still feasible when the domain size is less than or equal to 8 because the DP-summary construction requires only one execution.

Fig. 5 shows the construction time based on different bit sizes; this confirms that the DP-summary construction time increases linearly with the bit size representing ciphertexts. Thus, we confirm that our proposed method is feasible with a small domain size regardless of the bit size representing ciphertexts.

The DP-summary construction time is slow for two reasons. One is because of homomorphic operations. The computation cost over ciphertexts using homomorphic operations is large compared to that over plaintexts. In particular, TFHE is the bit-wise FHE, and the computation cost using TFHE is likely to be large compared to that using the integer-based FHE. The other is because any optimizations are impossible to be adapted. Original data-aware algorithm in DAWA [5] adopts some optimizations to speed up the partitioning. On the other hand, in our proposed method, any handled data is ciphertexts, which results in unavailability of any optimizations because we cannot see the values, i.e., plain texts, when encrypted. When partitioning, we need to seek the optimal partition from all possible partitions, whose computational complexity order is $O(2^n)$ ($n$ is domain size). Including these two reasons, it was reported that the computation using TFHE is approximately $10^9$ times slower than that over plaintexts [11].

During the measurement of the construction time, we also measured the maximum consumed memory size with the DP-command "bin/time --format=%M." The maximum consumed memory size varied from 513MB to 530MB depending on the domain size and the bit size.

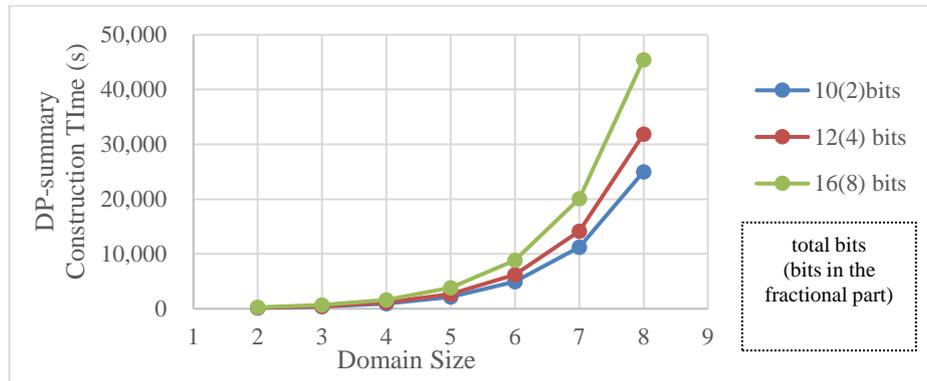

**Fig. 4.** DP-summary Construction Time v.s. Domain Size

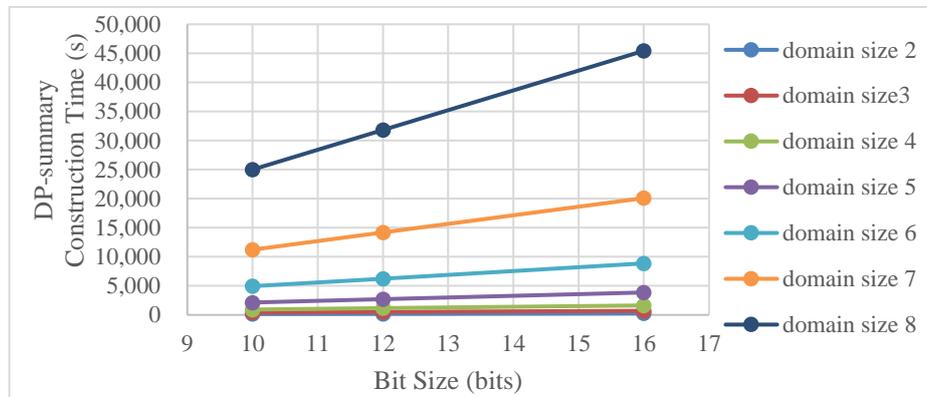

**Fig. 5.** DP-summary Construction Time v.s. Bit Size

### 5.3  Accuracy of DP-summary

In this experiment, we evaluated the accuracy of the DP-summary. In the implementation, we adopted a fixed-point number representation to truncate the fractional value according to the bit size representing the fractional part, which may affect the accuracy of the DP-summary. When the fractional part's bit size is small, the truncated fractional value becomes large, i.e., the accuracy of the DP-summary is expected to change depending on the fractional part's bit size. Thus, we measured the changes in the accuracy of the constructed DP-summary with different bit sizes of the fractional part.

We implemented the plaintext program that performs the same processing as our proposed method and the baseline using a floating-point number representation. This number representation is expressed in 64 bits: 1 bit for the code part, 52 bits for the mantissa part, and 11 bits for the exponent part.

We measured the error between the constructed DP-summary and the aggregated data on which differential privacy was not applied. We examined the error 100 times to determine the average for three different bit sizes—10(2) bits, 12(4) bits, and 16(8) bits—in a histogram with domain sizes from 2 to 10.

Fig. 6 shows the results of accuracy. We cannot verify the difference in accuracy according to the bit size. The reason is that, in our proposed method, the size of the truncated fractional part is negligibly small compared to the size of noise added by differential privacy; this implies that the effect on accuracy caused by changing the bit size of the fractional part is not significant.

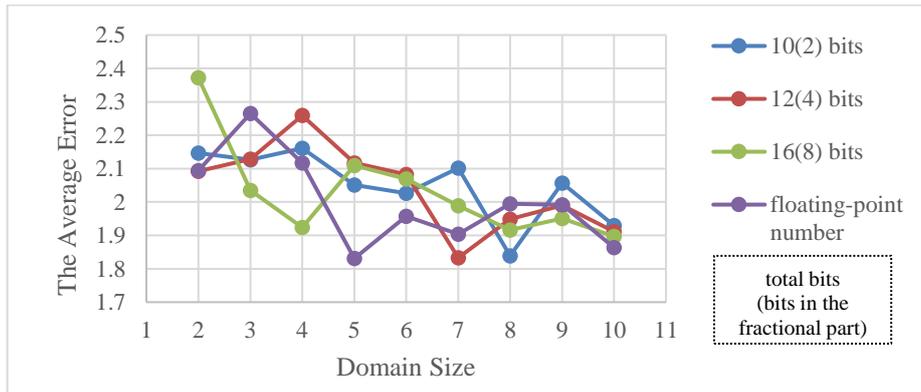

**Fig. 6.** Comparison of Accuracy v.s. Domain Size

## 6      Conclusion

In this study, we propose a privacy-aware query processing system that constructs differentially private summary (DP-summary) in advance over fully homomorphic encryption to respond to the queries using the DP-summary in plaintext. We enhance the data security in comparison with DP index proposed by Chowdhury et al. [4] in Crypt$\varepsilon$ system by applying differential privacy over the encrypted original data with FHE. Furthermore, we solve the problem of large added noise caused by partitioning the sorted data with equal -width regardless of the data distribution, by adopting the data-aware algorithm [5], which optimizes the partitioning depending on the data distribution.

Although our proposed method responds quickly to the queries, it requires high computational cost for DP-summary construction over fully homomorphic encryption. Our experimental evaluation with TFHE, which is a bit-wise fully homomorphic encryption library, shows that the proposed method requires 12.5 hours to construct DP-summary for eight 16-bits data; this is still feasible because the DP-summary requires only one construction. We also confirm that accuracy is not degraded even after adopting TFHE to the data-aware algorithm.

Our future work includes increasing the speed to prepare DP-summary for a larger dataset.

**Acknowledgment.** The research was supported by NII CRIS collaborative research program operated by NII CRIS and LINE Corporation.